# Hundred-thousand light holes push nanoscopy to go parallel

Xuanze Chen, Peng Xi*

Department of Biomedical Engineering, College of Engineering, Peking University, Beijing 100871, China

*Email: xipeng@pku.edu.cn

The recent work by Chmyrov et al. presents a capstone for the current major super-resolution microscopy techniques (Chmyrov et al. 2013). In optical super-resolution microscopy, two pathways are commonly taken: (1) targeted illumination modulation, such as stimulated emission depletion (STED) (Hell and Wichmann 1994), reversible/saturable optical fluorescence transition (RESOLFT) (Hell 2003), and saturated structured-illumination microscopy (SSIM) (Gustafsson 2005), or (2) stochastic modulation and single-molecule localization, such as (f)PALM/STORM (Betzig et al. 2006; Hess, Girirajan, and Mason 2006; Rust, Bates, and Zhuang 2006) in which the photo-switch nature of the fluorescent molecule is employed to achieve single-molecule imaging.

Of these methods, STED is more commonly regarded as an excitation point spread function being modulated by a doughnut shaped depletion PSF (Liu et al. 2012; Xie et al. 2013), although its name has no such implication. It is primarily the fact that stimulated emission requires a high intensity, and the doughnut is the practically most effective 2-D modulation pattern makes this impression. Consequently, STED is often referred as single-point sequential scanning as a typical confocal laser scanning microscopy (non-disk scan). The resolution of STED (or any type of doughnut modulated super-resolution technique) can be described with a square-root law: $d = d_0 / \sqrt{\xi}$ (Harke et al. 2008).

The limitation to single point scanning is often its imaging speed. As one can imagine, if *m* such focal points can be generated, then the image speed can be improved *m*× in principle through parallel imaging. Although efforts on generating multiple focal spots have been elaborated in STED (Bingen et al. 2011; Bingen 2012), the final result is often being trade-off by the quantity and quality of the focal "doughnuts", which is limited by the way the doughnut arrays are generated.

In 1803, Thomas Young demonstrated the wave nature of light through his famous double-slit, or double-hole interference. The same principle has been employed in structured illumination microscopy (SIM) (Lukosz 1966; Gustafsson 2000), in which the modulated illumination can be generated with two-beam interference in far field, originated from two pinholes to select the +1 and -1 diffraction orders, while blocking the central 0 order. Further decrease the thickness of the pattern can generate finer structures, either dark strips (in saturation case)(Gustafsson 2005) or bright stripes (in photo-switching case) (Hell 2003; Rego et al. 2012), which has been nicely illustrated in the explanation of the principle of SSIM in 1-D. The resolution enhancement of

SSIM is generally explained in frequency–domain with Moiré patterns and optical transfer function (OTF) extension (Heintzmann and Gustafsson 2009), in which multiple rotating angles and phases of periodical 1-D stripes have to be employed. In fact, such spatial-domain modulation with saturation/switching nature of the fluorescent dye has been clearly illustrated previously, and a uniformly square-root law has been given for standing wave modulation (Hell 2003). Moreover, spatial-domain explanation seems to be more straightforward in driving the question: how can this principle be extended in practical 2-D super-resolution imaging?

One immediate answer is, to sum two such orthogonal saturated fine patterns in 2-D. It is only recently that this elegant concept has been realized with incoherent crossed standing-waves (ICS) (Chmyrov et al. 2013). The pattern is generated similar to Young's two beam interference, but with four beams grouped in two orthogonal polarization states, resulting in a 2-D mesh (Fig. 1). Due to the fact that the sum of the two incoherent standing waves generates isotropic "doughnuts", the resulting effective excitation spots are homogenous and isotropic. Hence, a simple 2-D orthogonal scan can yield the final image; no frequency-domain post-processing is required as it can often lead to post-processing artifacts (Gustafsson, Agard, and Sedat 2000; Shroff, Fienup, and Williams 2010; Yang et al. 2014).

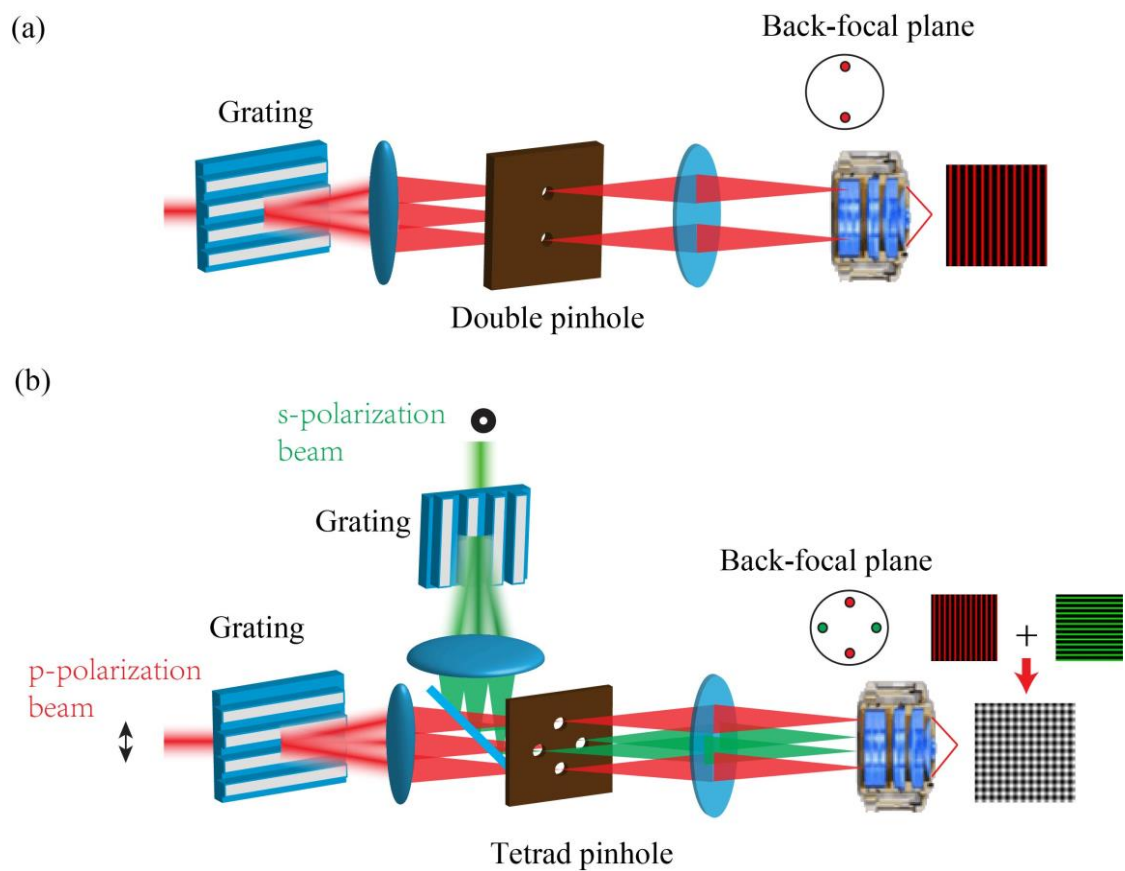

Fig. 1. (a) SIM uses interference to generate the 1-D illumination pattern. (b) In the parallel STED, a 2-D doughnut grid is generated with two orthogonal SIM

illumination. Red and green beams denote the p-polarization and s-polarization, respectively.

It should be noted that, in all kinds of super-resolution microscopy techniques, modulation plays the key role. In the conventional STED, the required modulation intensity is usually very high, thereby the application of multiple depletion doughnuts may be hindered by the physical threshold of the specimen and the fluorescent dye. In ICS, the on-off process through photoswitchable dye has been employed, instead of using stimulated emission. The application of photoswitchable dye for targeted modulation super-resolution was first raised in 2003 (Hell 2003) and then experimentally demonstrated in 2005 (Hofmann et al. 2005). Such application of photo-switchable dye due to their low saturation intensity has previously enabled 40nm resolution with RESLOFT spatial domain 2-D PSF narrowing (Grotjohann et al. 2011), and 50nm resolution with SSIM Fourier domain process (Rego et al. 2012). Notably, some earlier 1-D RESOLFT work already showed improved lateral resolution (Schwentker et al. 2007). The switching capability of the dye is also often used in localization super-resolution techniques. As this photo-switchable dye can be effectively turned off with light intensity 5 orders of magnitude lower than the saturation intensity required for STED, a 10-fold resolution improvement can be achieved with >1fps image acquisition speed.

Herein, ICS has utilized the concept of structured illumination in generating patterns, to generate effective 100,000 "doughnuts" as with STED; and used the photo-switchable dye to decrease the requirement of modulation intensity. It has combined the key elements of all these major super-resolution techniques, with a single aim: faster and better resolution for dynamic biological study. It is worth noting that, ICS/RESOLFT/SSIM employed spatial saturation of the fluorescence excitation to obtain super-resolution. SSIM explained the resolution enhancement with higher harmonic generation. Similarly, the saturation-harmonics generation can be modulated and detected in time-domain (Fujita et al. 2007; Yamanaka et al. 2013), which may open up a new horizon for current spatial modulation approaches.

So far, it seems that ICS has made a new summit of optical super-resolution microscopy. What may come next? In Chmyrov et al.'s work, the use of the photo-switchable fluorescence protein may limit some of the biological application. Alternatively, ground state depletion (GSD) has already enabled common organic fluorescent dyes with photo-switching capability (Hell and Kroug 1995; Bretschneider, Eggeling, and Hell 2007; Fölling et al. 2008). Since the requirement of photoswitchable dye is for its low modulation intensity, and many novel alternative inorganic dyes can play this role (Irvine et al. 2008; Han et al. 2010; Kolesov et al. 2011; Díaz et al. 2011), a wide-spread of the ICS parallel nanoscopy technique, in conjugation with the commonly used organic fluorophores/inorganic counterparts, may be readily foreseeable in the near future, toward higher resolution and more dynamic live cell imaging.

**Acknowledgements**

This work is supported by the National Instrumentation Program (2013YQ03065102), the "973" Major State Basic Research Development Program of China (2011CB809101, 2010CB933901), and the National Natural Science Foundation of China (61178076).